\documentclass[11pt]{article}
\usepackage{graphics,graphicx,dcolumn,bm,epic,eepic,float}
\usepackage{amssymb,multirow,rotate,color}

\def \bsigma {\mbox {\boldmath $\sigma$}}
\def \bxi {\mbox {\boldmath $\xi$}}
\def \bm {\mbox {\boldmath $m$}}
\def \btm {\mbox {\boldmath $\tilde m$}}

\def \bI {\mbox {\boldmath $I$}}
\def \bq {\mbox {\boldmath $q$}}
\def \bDz {\mbox {\boldmath $Dz$}}
\def \bz {\mbox {\boldmath $z$}}

\begin{document}

\setcounter{page}{0}

\title{\bf Feed-forward chains of recurrent attractor neural networks
with finite dilution near saturation}
\author{F. L. Metz and W. K. Theumann \footnote{e-mail: theumann@if.ufrgs.br\,\,\,\,
phone: +55-51-3316.6486\,\,\,\,fax:+55-51-3316.7286}\\
Instituto de F\'{\i}sica, Universidade Federal do Rio Grande do
Sul,\\ Caixa Postal 15051. 91501-970 Porto Alegre, RS, Brazil.}

\date{\today}
\maketitle
\thispagestyle{empty}

\begin{abstract}

A stationary state replica analysis for a dual neural network
model that interpolates between a fully recurrent symmetric
attractor network and a strictly feed-forward layered network,
studied by Coolen and Viana, is extended in this work to account
for finite dilution of the recurrent Hebbian interactions between
binary Ising units within each layer. Gradual dilution is found to
suppress part of the phase transitions that arise from the
competition between recurrent and feed-forward operation modes of
the network. Despite that, a long chain of layers still exhibits a
relatively good performance under finite dilution for a balanced
ratio between inter-layer and intra-layer interactions. \\
\mbox{} \\
\indent Pacs: 87.10.+e, 64.60.Cn, 07.05.Mh

\end{abstract}

\vspace{0.5cm} \setcounter{page}{1}

\section{Introduction}

It is now well established that biological and artificial neural
networks propagate information by means of synaptic interactions
which may be symmetric or non-symmetric. The latter appear in
biological networks, while either of both have been used in
artificial networks.

There are two main classes of neural network models for
associative information processing. One is that of recurrent
networks with feed-back of either symmetric or non-symmetric
synapses \cite{Ho82,HKP91} and the other one is that of layered
feed-forward networks, with no feed-back loops, in which the
synapses are intrinsically non-symmetric
\cite{DMK86}-\cite{DKM89}. Models of the first class can be solved
by means of equilibrium statistical mechanics in the case of
symmetric synapses \cite{AGS85,AGS87}, due to the presence of
detailed balance, whereas one has to resort to a dynamical
procedure in the search for stationary states in the case of
non-symmetric synapses \cite{Co01}. On the other hand,
feed-forward networks are already dynamical systems which may
reach stationary states in place of true equilibrium states.

A novel dual model of binary neurons which combines recurrent and
feed-forward processing, with symmetric synapses in the recurrent
part, has been proposed some time ago by Coolen and Viana
\cite{CV96}, and the model has been generalized recently
\cite{KH01}-\cite{KSH03}. The network model consists of layers of
fully connected neurons in a recurrent architecture that receive
inputs from the previous layer (except the first one) and pass the
processed information by means of feed-forward synaptic
interactions to the next layer. The latter processes further the
information and feeds, in turn, the next layer. Every unit in each
layer is assumed to be connected to every other unit in that layer
as well as to every unit in the next and in the previous layer but
the information goes only from one layer to the next one (the
feed-forward nature of the model). It is important to note that
each unit in every layer participates of both the recurrent and
the feed-forward process. Thus, a single updating procedure
(usually, either sequential or parallel) has to be applied to
every unit in each layer.

The dual model becomes solvable by means of equilibrium
statistical mechanics when all layers reach a stationary state and
that may eventually be the case in the zero-temperature limit. In
the case of finite temperature, thermal fluctuations destroy the
Boltzmann form of the stationary states in the layers and one has
to go over to a dynamical approach, which has not been done
apparently so far. Actually, the model has been solved working out
the free energy and the equations for the order parameters for
finite temperature, assuming a Boltzmann form based on an
underlying sequential updating dynamics, and letting the
temperature go to zero after the thermodynamic limit. The close
agreement between theory and numerical simulations justified the
procedure and the order of these limits despite the absence of
detailed balance \cite{CV96}.

A feed-forward chain of recurrent networks is interesting as a
model of dual information processing in which the feed-forward
mode transmits the outcome for the stationary overlaps within a
layer to the next one. So far, only the case of a fully connected
network has been considered, and one may ask how does the
performance of the model change if the constraint of full
connectivity between the units within each layer is relaxed by
means of synaptic dilution keeping the full connectivity between
units in adjacent layers. Synaptic dilution reduces the outcome
from a layer and it could be that the cumulative effect in a long
chain of layers would be a vanishingly small storage capacity even
for a reduced amount of dilution.

This issue is of interest in statistical mechanics and also for
the information processing in biological networks, in particular
for the understanding of the role played by the $CA_3$ region of
the hippocampus in the primate brain where the mean connection
between units (our parameter $c$ below) in the latter is of the
order of $0.1 - 0.01$ \cite{RTP95,SR99}, which may be considered
as a finite dilution. Finite dilution may also be of interest if
one thinks of defects in the connectivity within layers in
artificial neural networks. Detailed studies of the dynamics and
the equilibrium states that describe the performance have been
done in recent works that deal with finite dilution in various
neural network models on different architectures
\cite{KT99}-\cite{BEV05} following earlier works
\cite{So86}-\cite{CN92}.

The main purpose of the present paper is to study the stationary
states and obtain the phase diagrams that describe the performance
of the dual model with finite dilution of the synaptic
interactions within layers. In consistency with the procedure, we
first take into account the disorder due to dilution for non-zero
temperature, solve the model in the thermodynamic limit and then
let the temperature go to zero. A second, minor purpose, is to
indicate the necessary steps that yield the recursion relation for
the parameter that accounts for the input from the non-condensed
overlaps which follows an unconventional derivation. It will be
shown that, in addition to the well-known effect of finite
dilution in replacing the intra-layer synaptic connection by an
effective interaction that is the superposition of a Hebbian term
for a fully connected network and a Gaussian noise, the only
further contribution of the dilution on that parameter is an
overall factor which can be absorbed by a simple rescaling in the
case of dense networks. A further interest in the procedure to
deal with the effects of synaptic dilution is that it may be
extended to study the influence of general non-linear synapses
\cite{So86}.

The outline of the paper is the following. In Sec. 2 we review the
model and introduce the alterations due to finite dilution. In
Sec. 3 we solve the model to obtain the replica symmetric free
energy near saturation and derive the recursion relations that
describe the evolution of the parameters of the network from one
layer to the next. The phase transitions and some features of the
performance are obtained and discussed in Sec. 4 and we end with
concluding remarks in Sec. 5.

\section{The model}

The network model consists of $L$ layers with $N$ binary Ising
units (neurons) on each layer $l$ in a microscopic state
$\bsigma^l=\{\sigma_1^l,\dots,\sigma_N^l\}$, in which each
$\sigma_i^l=\pm 1$. The state $+1$ represents a firing neuron and
the state $-1$ a neuron at rest. The microscopic dynamics of the
network is assumed to be a Glauber sequential stochastic alignment
of each neuron $\sigma_i^l$ to a local field $h_i^l$ with
probability
\begin{equation}
{\rm Prob}(\sigma_i^{l}\rightarrow -\sigma_i^{l})
=\frac{1}{2}[1-\tanh(\beta\sigma_i^{l} h_i^{l})]\,\,\,\,,
\label{1}
\end{equation}
where
\begin{equation}
h_i^l(\bsigma^l,\bsigma^{l-1})=\sum^{N}_{j=1}J_{ij}^l\sigma_j^l +
\sum^{N}_{j=1}K_{ij}^l\sigma_j^{l-1}\,\,\,\,, \label{2}
\end{equation}
is due to the states of other (see below) neurons $\bsigma^l$ on
the same layer and of neurons $\bsigma^{l-1}$ on the previous
layer, as shown schematically in Fig.1. At each time step the
neuron to be updated is taken at random from the set
$\{\bsigma^l\}$. The parameter $\beta=T^{-1}$ controls the
synaptic noise such that the dynamics of the network becomes
deterministic when $T\rightarrow 0$.

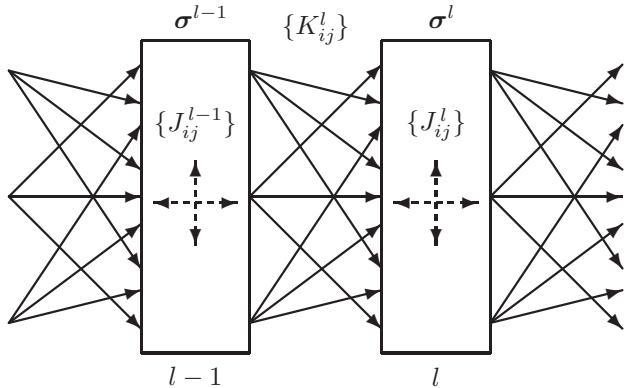
\begin{figure}[H]
\begin{center}
\setlength{\unitlength}{0.8mm}

\begin{picture}(140,60)(-20,0)
\thicklines

\put(22,0){\line(0,1){52}}
\put(40,0){\line(0,1){52}}
\put(22,0){\line(1,0){18}}
\put(22,52){\line(1,0){18}}

\put(62,0){\line(0,1){52}}
\put(80,0){\line(0,1){52}}
\put(62,0){\line(1,0){18}}
\put(62,52){\line(1,0){18}}

\put(0,5){\vector(4,1){22}}
\put(0,5){\vector(4,3){22}}
\put(0,5){\vector(2,3){22}}

\put(0,26){\vector(1,0){22}}
\put(0,26){\vector(1,1){22}}
\put(0,26){\vector(1,-1){22}}

\put(0,47){\vector(4,-1){22}}
\put(0,47){\vector(4,-3){22}}
\put(0,47){\vector(2,-3){22}}

\put(40,5){\vector(4,1){22}}
\put(40,5){\vector(4,3){22}}
\put(40,5){\vector(2,3){22}}

\put(40,26){\vector(1,0){22}}
\put(40,26){\vector(1,1){22}}
\put(40,26){\vector(1,-1){22}}

\put(40,47){\vector(4,-1){22}}
\put(40,47){\vector(4,-3){22}}
\put(40,47){\vector(2,-3){22}}

\put(80,5){\vector(4,1){22}}
\put(80,5){\vector(4,3){22}}
\put(80,5){\vector(2,3){22}}

\put(80,26){\vector(1,0){22}}
\put(80,26){\vector(1,1){22}}
\put(80,26){\vector(1,-1){22}}

\put(80,47){\vector(4,-1){22}}
\put(80,47){\vector(4,-3){22}}
\put(80,47){\vector(2,-3){22}}

\put(31,-4){\makebox(0,0)[c]{\small{$l-1$}}}
\put(71,-4){\makebox(0,0)[c]{\small{$l$}}}
\put(72,56){\makebox(0,0)[c]{\small{$\bsigma^{l}$}}}
\put(32,56){\makebox(0,0)[c]{\small{$\bsigma^{l-1}$}}}
\put(51,54){\makebox(0,0)[c]{\small{$\{K_{ij}^{l}$\}}}}
\put(31,38){\makebox(0,0)[c]{\small{$\{J_{ij}^{\,l-1}$\}}}}
\put(71,38){\makebox(0,0)[c]{\small{$\{J_{ij}^{\,l}$\}}}}

\put(25,25){\vector(-1,0){1}}
\put(27,25){\line(1,0){1}}
\put(29,25){\line(1,0){1}}
\put(31,25){\line(1,0){1}}
\put(33,25){\line(1,0){1}}
\put(35,25){\line(1,0){1}}
\put(37,25){\vector(1,0){1}}

\put(31,31){\vector(0,1){1}}
\put(31,29){\line(0,-1){1}}
\put(31,27){\line(0,-1){1}}
\put(31,25){\line(0,-1){1}}
\put(31,23){\line(0,-1){1}}
\put(31,21){\line(0,-1){1}}
\put(31,19){\vector(0,-1){1}}

\put(65,25){\vector(-1,0){1}}
\put(67,25){\line(1,0){1}}
\put(69,25){\line(1,0){1}}
\put(71,25){\line(1,0){1}}
\put(73,25){\line(1,0){1}}
\put(75,25){\line(1,0){1}}
\put(77,25){\vector(1,0){1}}

\put(71,31){\vector(0,1){1}}
\put(71,29){\line(0,-1){1}}
\put(71,27){\line(0,-1){1}}
\put(71,25){\line(0,-1){1}}
\put(71,23){\line(0,-1){1}}
\put(71,21){\line(0,-1){1}}
\put(71,19){\vector(0,-1){1}}

\end{picture}

\end{center}
\caption{\protect  The model. Two consecutive layers in a chain of
length $L$ with neuron states $\bsigma^l$ and symmetrically
diluted recurrent interactions $\{J_{ij}^l\}$ within layer $l$ and
feed-forward interactions $\{K_{ij}^l\}$ from the previous layer.}
\label{fig1}
\end{figure}

A macroscopic set of $p=\alpha N$ independent and identically
distributed random patterns $\bxi^{\mu,l}
=\{\xi_1^{\mu,l},\dots,\xi_N^{\mu,l}\}$; $\mu=1,\dots,p$, are
stored on the sites of each layer in the learning stage following
a Hebbian rule
\begin{equation}
J_{ij}^l=\frac{c_{ij}J_0}{cN}(1-\delta_{ij})\sum_{\mu=1}^p\xi_{i}^{\mu,l}
\xi_{j}^{\mu,l}\,\,\,\,\,;\,\,\,\,\,
K_{ij}^l=\frac{J}{N}\sum_{\mu=1}^p\xi_{i}^{\mu,l}\xi_{j}^{\mu,{l-1}}
\label{3}
\end{equation}
where each $\xi_i^{\mu,l}=\pm 1$, with probability $\frac{1}{2}$,
is the component of the pattern $\mu$ on neuron $i$ of layer $l$.
Thus, $\alpha=p/N$ is the storage ratio of patterns {\it{per
site}} of the Hopfield model \cite{Ho82}, whether the sites are
connected or not within a layer, since they are always connected
to the sites of the adjacent layers. The parameters $(J_0, J)$
control the relative strength of the recurrent and feed-forward
interactions, respectively, and they are the same for all layers,
except on the first layer where $K_{ij}^1=0$, for all $(i,j)$, and
where one may distinguish between a free relaxation and a clamped
operation defined below \cite{CV96}. The recurrent part of the
synaptic interaction is symmetrically diluted by means of the set
of identically distributed random variables $\{c_{ij}\}$, with
$c_{ij}=c_{ji}$, such that $c_{ij}=1$ with probability $c$ and
zero with probability $1-c$. The average number of connected
neurons in each layer, the so-called connectivity, is $cN$ and we
focus our attention in this work on dense networks in which
$c=O(1)$ implying that $cN \gg 1$ within each layer \cite{So86}.
When $c=1$ the network is fully connected and letting $c$ go to
zero, after the thermodynamic limit $N \rightarrow\infty$, one has
the network with strongly (symmetrical) diluted layers. One should
expect that the synaptic dilution within the layers will have
effects on the feed-forward information processing of the network,
as long as $J_0\neq 0$, and it is this that we want to study here.

When $J=0$ the model reduces to a set of $L$ decoupled
symmetrically diluted attractor Hopfield-like networks which can
be solved by means of equilibrium statistical mechanics using the
replica method \cite{So86,WS91,CN92}. In that case, the
intra-layer synaptic interaction becomes an effective interaction
which is the sum of a Hebbian term and a Gaussian noise that will
be determined below. In the case of full connectivity ($c=1$) the
noise term vanishes and the critical storage ratio is that of the
usual Hopfield model, $\alpha_c \approx 0.138$.

In contrast, when $J_0=0$, we have a purely feed-forward network
with full inter-layer synaptic interactions between neurons on
consecutive layers which can be solved exactly for all $\alpha$,
including the saturation of stored patterns, in the large $N$
limit by means of a signal-to-noise analysis which is a truncated
version of a neurodynamic approach developed some time ago
\cite{AM88,Ok96}. In that case, the local fields follow a Gaussian
distribution allowing for the derivation of exact recursion
relations for the order parameters in the limit of
$N\rightarrow\infty$ \cite{DKM89} with a critical storage capacity
$\alpha_c \approx 0.269$. We expect this capacity, for $J_0=0$, to
remain the same under exclusive synaptic dilution within the
layers but there should be a gradual change of $\alpha_c$ with
increasing $J_0$ and an eventual maximum $\alpha_c$ for an
intermediate situation between feed-forward and recurrent
operation, as in the case of the fully connected network ($c=1$)
where $\alpha_c \approx 0.317$ \cite{CV96}.

Coming back to the input at the first layer, one may have either a
free relaxation from a specific initialization as a recall cue or
a so-called "clamped" operation where the recall cue is given by
an externally specified but stationary random state vector on that
layer. We will discuss the effects of dilution on both modes.

Due to the feed-forward nature of the interactions between
consecutive layers the network will attain a stationary state at
$T=0$. This will be the case for any layer $l$ when the input from
layer $l-1$ becomes stationary. The equilibrium states of the
network may then be associated with the stationary states of the
dynamics with a probability distribution given by a Boltzmann form
$p(\bsigma^l)\sim e^{-\beta H(\bsigma^l)}$, for any layer, with
the Hamiltonian
\begin{equation}
H(\bsigma^l)=-\frac{1}{2}\sum_{ij}J_{ij}^l\sigma_i^l\sigma_j^l-
\sum_{ij}K_{ij}^l\sigma_i^l\sigma_j^{l-1}\,\,\,\,,
\label{4}
\end{equation}
which enables one to solve the model first for finite $T$ and then
taking the $T \rightarrow 0$ limit, as in previous work
\cite{CV96}.

To study the performance of the network we introduce the
macroscopic overlaps between the state and the stored patterns on
that layer,
\begin{equation}
m_{\mu,l}(\bsigma^l)=\frac{1}{N}\sum_i\xi_i^{\mu,l}\sigma_i^l
\label{5}
\end{equation}
and assume, as usual, that in equilibrium only a finite number $k$
of the patterns is condensed with overlap of $O(1)$, for
$\mu=1,\dots,k$, and the remaining $p-k$ overlaps are assumed to
be macroscopically small of $O(1/\sqrt N)$. The local field
$h_i^l(\bsigma^l,\bsigma^{l-1})$ will not only have contributions
from these overlaps but also from the stationary overlaps of the
previous layer $l-1$, due to the second term in Eq.(2),
\begin{equation}
\tilde
m_{\mu}=\frac{1}{N}\sum_i\xi_i^{\mu,l-1}\sigma_i^{l-1}\,\,\,\,,
\label{6}
\end{equation}
with similarly defined condensed and non-condensed components.

\section{Free energy and stationary states}

We consider the partition function
\begin{equation}
Z=\sum_{\bsigma}e^{-\beta H(\bsigma)}\,\,\,\,, \label{7}
\end{equation}
with $H(\bsigma)$ given by Eq.(4), and calculate the free energy
treating, as usual, the non-condensed patterns $\mu>k$ as quenched
disorder over which we average the logarithm of the partition
function by means of the replica method to obtain the free energy
$f$ per site as
\begin{equation}
-\beta f=\lim_{n\rightarrow 0}\lim_{cN \rightarrow\infty}
\frac{1}{Nn} \{ \langle[Z^n]_c\rangle_{patt}-1 \} \,\,\,\,. \label{8}
\end{equation}
Here, $\langle[Z^n]_c\rangle_{patt}$ is the average over the
quenched disorder, due to the patterns, of the connected
replicated partition function $[Z^n]_c$ obtained by taking the
average over the random symmetrical dilution.

With the dilution only of synapses within the layers, we may write
\begin{equation}
[Z^n]_c=\sum_{\{\bsigma^1\}\dots\{\bsigma^n\}}e^{\beta\sum_{\alpha}\sum_{ij}
K_{ij}^l\sigma_{i\alpha}^l\sigma_{j\alpha}^{l-1}}\prod_{i<j}\big[e^{\beta
J_{ij}^l\sum_{\alpha}\sigma_{i\alpha}^l\sigma_{j\alpha}^l}\big]_c\,\,\,\,,
 \label{9}
\end{equation}
where $\alpha$ (and $\beta$, below) denotes the replica index,
whenever it appears as a subindex, and $\{\bsigma^n\}$ is the set
of state vectors in the replica $n$. Using the now standard
procedure to perform the configurational average over the
symmetric dilution in the case of weak synapses
$J_{ij}^l=O(\frac{J_0}{c}\sqrt{\frac{\alpha}{N}})$ in the large
$N$ limit one finds
\begin{eqnarray}
\prod_{i<j}\big[\exp(\beta
J_{ij}^l\sum_{\alpha}\sigma_{i\alpha}^l\sigma_{j\alpha}^l)\big]_c&=&
\exp\Big\{\frac{\beta^2J_0^2\alpha
(1-c)}{4cN}\sum_{\alpha\beta}(\sum_i\sigma_{i\alpha}^l\sigma_{i\beta}^l)^2\Big\}
\nonumber\\
&\times &\exp\Big\{\frac{\beta NJ_0}{2}\sum_{\mu=1}^{p}
\sum_{\alpha}m_{\mu,l}^2(\bsigma_{\alpha})\Big\}\,\,\,\,. \label
{10}
\end{eqnarray}
Thus, the effect of the dilution amounts to the replacement of the
intralayer synapses $J_{ij}^l$ by the effective interaction
\cite{So86,CN92}
\begin{equation}
J_{ij\,\,eff}^{l}=\frac{J_0}{N}\sum_{\mu=1}^p\xi_{i}^{\mu,l}
\xi_{j}^{\mu,l}+\delta_{ij}^l\,\,\,\,, \label{11}
\end{equation}
for $i\neq j$ which is the sum of a Hebbian term for a fully
connected recurrent network and a Gaussian noise $\delta_{ij}^l$
of mean zero and variance
\begin{equation}
\Delta^2=[(\delta_{ij}^l)^2]_c=J_0^2\alpha (1-c)/c\,\,\,\,.
\label{12}
\end{equation}

Coming back to the averaged partition function we obtain

\begin{eqnarray}
\langle[Z^n]_c\rangle_{patt}&=&e^{-\frac{1}{2}J_0\alpha\beta
nN}\nonumber\\
&\times&\Big\langle\sum_{\{\bsigma^1\}\dots\{\bsigma^n\}}
\exp{\Big\{\frac{\beta^2J_0^2\alpha(1-c)}{4cN}\sum_{\alpha\beta}
(\sum_i\sigma_{i\alpha}^l\sigma_{i\beta}^l)^2\Big\}}\nonumber\\
&\times&\exp{\Big\{\beta
N\sum_{\mu}\sum_{\alpha}\big[\frac{J_0}{2}m_{\mu,l}^2(\bsigma_{\alpha}^l)
+Jm_{\mu,l}(\bsigma_{\alpha}^l)\tilde
m_{\mu}\big]}\Big\}\Big\rangle_{\bxi}\,\,\,\, \label{13}
\end{eqnarray}
where the average is over the patterns $\bxi$ in the overlaps
$m_{\mu,l}(\bsigma_{\alpha}^l)$, and $\tilde m_{\mu}$ are the
already stationary overlaps in the previous layer. The sum over
$\mu$ in the exponent is over both, $\mu\leq k$ (condensed) and
$\mu> k$ (non-condensed) patterns.

We follow the extension of the standard procedure \cite{CV96} to
calculate $\langle[Z^n]_c\rangle_{patt}$ and obtain the free
energy per site in the large-$N$ limit by means of the method of
steepest descent. We find the replica-symmetric form
\begin{eqnarray}
f&=&\frac{1}{2}\alpha [J_0+\beta
r(1-q)]+\frac{1}{2}J_0\sum_{\mu\leq k}m_{\mu}^2 +
\frac{\alpha}{2\beta}\Big\{\log[1-\beta
J_0(1-q)]\nonumber\\
&-&\frac{J_0\beta q}{1-\beta J_0(1-q)}\Big\} -\frac{\beta
J^2(1-q)}{2\big[1-\beta J_0(1-q)\big]}\sum_{\mu >k}\tilde
m_{\mu}^2-\frac{1}{4}\beta \Delta^2(1-q^2)\nonumber\\
&-&\frac{1}{\beta}\Big\langle\int Dz \log\Big\{
2\cosh\beta\big[\sum_{\mu\leq k} \xi_{\mu}(J_0m_{\mu}+J\tilde
m_{\mu})+z\sqrt{\alpha r}\big]\Big\}\Big\rangle_{\{\xi_{\mu}\}}\,,
\label{14}
\end{eqnarray}
where $\Delta^2$ is given by Eq.(12),
$Dz=(2\pi)^{-\frac{1}{2}}e^{-\frac{1}{2}z^2}dz$ and all quantities
refer to any given layer, except for $\tilde m_{\mu}$ which refers
to the previous layer. Here, $m_{\mu}$, $q$ and $r$ are the
replica-symmetric parameters that will be determined as solutions
of the saddle-point equations $\partial f/\partial
m_{\mu}=\partial f/\partial q =\partial f/\partial r=0$. The
condensed overlap $\tilde m_{\mu}$ for $\mu \leq k$ will be taken
as an input from the previous layer and the sum over non-condensed
patterns, $\sum_{\mu>k}\tilde m_{\mu}^2$, provides a further link
with that layer which will be determined below.

The parameters are interpreted as follows: $m_{\mu}=\langle
m_{\mu,l}(\bsigma_{\alpha}^l)\rangle$ is the replica symmetric
average overlap with pattern $\bxi^{\mu l}$ for all replicas
$\alpha$ given by Eq.(5), while
$q=\langle\sigma_{i\alpha}^l\sigma_{i\beta}^l\rangle$ is the
replica symmetric average of the spin-glass order parameter
$q_{\alpha\beta}
=\frac{1}{N}\sum_i\sigma_{i\alpha}^l\sigma_{i\beta}^l$, for
$\alpha\neq\beta$, where the brackets stand for a thermal and
configurational average over the patterns. As usual, $r$ accounts
for the introduction of the spin-glass order parameter into the
replicated partition function and it is given here by
\begin{eqnarray}
\alpha\beta^2r&=&\beta^2J^2x_{\alpha}x_{\beta}\sum_{\mu>k}\tilde
m_{\mu}^2+\beta^2\Delta^2 q\nonumber\\
&+&\alpha\beta
J_0\frac{\int \bDz\,z_{\alpha}z_{\beta}\,e^{-\frac{1}{2}\beta
J_0\bz.\bq\bz}} {\bDz\,z\,e^{-\frac{1}{2}\beta J_0\bz.\bq\bz}}
\,\,\,\,, \label{15}
\end{eqnarray}
for $\alpha\neq\beta$, where $x_{\alpha}=\sum_{\gamma}[\bI-\beta
J_0\bq]_{\alpha\gamma}^{-1}$ in which $\bI$ is the identity matrix
and $\bq=\{q_{\alpha\beta}\}$, while $\bz\in\Re^{n}$ and
$\bDz=(2\pi)^{-n/2}e^{-\frac{1}{2}z^2}d\bz$.

In order to determine $\sum_{\mu>k}\tilde m_{\mu}^2$, consider the
formal relationship for the derivative of the free energy with
respect to the control parameter $J_0$, which can be derived
directly from Eqs.(8), (12) and (13)
\begin{equation}
\sum_{\mu >k}\langle m_{\mu,l}^2(\bsigma^l)\rangle =\alpha
-\alpha\beta J_0\frac{1-c}{c}(1-q^2) -\sum_{\mu\leq k}\langle
m_{\mu,l}^2(\bsigma^l)\rangle-2\frac{\partial f}{\partial
J_0}\,\,\,\,, \label{16}
\end{equation}
in replica symmetric form. Note that the second term on the
right-hand side contains the noise in the form $-\beta
(1-q^2)\Delta^2/J_0$ that will be cancelled by a similar term from
the last part of the equation. Using the explicit expression for
the free energy in Eq.(14), one finds
\begin{eqnarray}
\sum_{\mu >k}\langle m_{\mu,l}^2(\bsigma^l)\rangle
&=&\frac{1}{[1-\beta
J_0(1-q)]^2} \{\alpha [1-\beta J_0(1-q)^2]\nonumber\\
&+&\beta^2 J^2(1-q)^2 \sum_{\mu>k}\langle
m_{\mu,l-1}^2(\bsigma^{l-1})\rangle\}\,\,\,\,, \label{17}
\end{eqnarray}
relating the non-condensed overlaps in two consecutive layers,
where $q$ is the spin-glass order parameter in layer $l$. A
similar relationship can be derived for layer $l-1$ where $\langle
m_{\mu,l-2}^2\rangle$ becomes the stationary input $\tilde
m_{\mu}^2$ for that layer. Using this relationship with the
saddle-point equation $\partial f/\partial q=0$ for layer $l-1$,
in order to eliminate $\sum_{\mu>k}\langle m_{\mu,l-2}^2\rangle$,
and applying again the saddle-point equation to layer $l$ we
obtain the third recursion relation below between the macroscopic
order parameters in two consecutive layers $(\bm, q, r)\rightarrow
(\bm^{\prime}, q^{\prime}, r^{\prime})$.

Thus, altogether, the saddle-point equations for the dense network
in which $c=O(1)$ yield the recursion relations
\begin{equation}
\bm^{\prime}=\Big\langle\bxi\int Dz \tanh
\beta\Big\{\bxi.(J_0\bm^{\prime} +J\bm)+z\sqrt{\alpha [\tilde
r^{\prime}+J_0^2(1-c)q^{\prime}/c]}\,\Big\}\Big\rangle_{\bxi}
\label{18}
\end{equation}
\begin{equation}
q^{\prime}=\Big\langle\int Dz \tanh^{2}
\beta\Big\{\bxi.(J_0\bm^{\prime} +J\bm)+z\sqrt{\alpha [\tilde
r^{\prime}+J_0^2(1-c)q^{\prime}/c]}\,\Big\}\Big\rangle_{\bxi}
\label{19}
\end{equation}
\begin{equation}
\tilde r^{\prime}[1-\beta J_0(1-q^{\prime})]^2-J_0^2q^{\prime}=
\beta^2 J^2(1-q)^2\tilde r-J^2q+\frac{J^2(1+q)}{1-\beta
J_0(1-q)}\,\,\,\,, \label{20}
\end{equation}\\
where the last equation is formally the same as that for the fully
connected recurrent layers in terms of the rescaled variable
$\tilde r=r/c$. Here, $\bm=(m_1,\dots,m_k)$ is the condensed
overlap vector in one layer and
$\bm^{\prime}=(m^{\prime}_1,\dots,m^{\prime}_k)$ in the next
layer. Given a state of the first layer, Eqs.(18)-(20) describe
the evolution of the states of the network. We recover the
fixed-point equations for a diluted purely recurrent network when
$J=0$ for $J_0=1$ \cite{CN92} and the recursion relations for the
layered feed-forward network, with full connection of units
between layers, when $J_0=0$ for $J=1$ \cite{DKM89}. For $c=1$ we
recover the equations of Coolen and Viana \cite{CV96}.

Note that the synaptic dilution in the case of the dense network
enters only through the variance of the noise $\Delta^2$ in the
effective intra-layer interactions. In contrast, in the extremely
dilute limit $c\rightarrow 0$, it is more appropriate to rescale
\cite{WS91,CN92}
$\alpha\,\rightarrow\alpha^{\prime}=\alpha/c=p/cN$, the ratio of
stored patterns per mean number of {\it{connected}} sites, in
place of rescaling $r$. Eq.(20) becomes then a recursion relation
that depends explicitly on $c$.

Following earlier work, we consider either a clamped input or a
free relaxation in the first layer \cite{CV96}. In the first case,
the state $\bsigma^1$ in that layer may be any randomly chosen
configuration with a given condensed overlap $\bm$. The first two
recursion relations remain the same and the last one is replaced
by
\begin{equation}
\tilde r^{\prime}[1-\beta
J_0(1-q^{\prime})]^2-J_0^2q^{\prime}=J^2\,\,\,\,. \label{21}
\end{equation} 
On the other hand, in the case of a free relaxation the
macroscopic state $(\bm, q, r)$ of the first layer that follows
from Eqs.(18)-(20) when $J=0$ is determined by the equations
\begin{eqnarray}
\bm&=&\Big\langle\bxi\int Dz \tanh \beta\Big\{\bxi.J_0\bm
+z\sqrt{\alpha
[\tilde r+J_0^2(1-c)q/c]}\,\Big\}\Big\rangle_{\bxi}\\
q&=&\Big\langle\int Dz \tanh^{2} \beta\Big\{\bxi.J_0\bm
+z\sqrt{\alpha
[\tilde r+J_0^2(1-c)q/c]}\,\Big\}\Big\rangle_{\bxi}\\
\tilde r&=&\frac{J_0^2q}{[1-\beta J_0(1-q)]^2} \,\,\,\,.
\label{24}
\end{eqnarray}\\
We recover the known equations for the diluted recurrent network
when $J_0=1$ \cite{CN92}. Thus, Eqs.(18)-(24) extend the model of
Coolen and Viana for the case of symmetric synaptic dilution in
the recurrent layers. We take next the limit $T\rightarrow 0$ and
eliminate the parameter redundancy writing
$J_0=\frac{1}{2}(1+\omega)$ and $J=\frac{1}{2}(1-\omega)$
\cite{CV96}.

\section{Phase transitions}

We turn next to the effects of dilution on the phase transitions
in the model and consider three situations as in earlier work
\cite{CV96}. First, we look for the storage capacity of infinitely
long chains in which a stationary state is reached. Then we focus
attention on the first and the second layer in a search for
multiple transitions between replica-symmetric states. The third
transition we discuss is the replica-symmetry breaking transition
at the de Almeida-Thouless (AT) line.\\
\mbox{} \\
\noindent 4.1 {\it{Saturation transition in an infinitely long
diluted chain}}
\\
\mbox{} \\
We consider here the transition where the overlap disappears with
an increasingly large storage of patterns. A stationary state is
reached along the network when $(\bm^{\prime}, q^{\prime}, \tilde
r^{\prime})=(\bm, q, \tilde r)$ and Eqs.(18)-(20) become

\begin{eqnarray}
\bm&=&\Big\langle\bxi\int Dz \tanh \beta\Big\{\bxi.\bm
+z\sqrt{\alpha
[\tilde r+J_0^2(1-c)q/c]}\,\Big\}\Big\rangle_{\bxi}\\
q&=&\Big\langle\int Dz \tanh^{2} \beta\Big\{\bxi.\bm
+z\sqrt{\alpha
[\tilde r+J_0^2(1-c)q/c]}\,\Big\}\Big\rangle_{\bxi}\\
\tilde r&=&\frac{(1-\omega)^2+q(1+\omega)^2-2qC\omega (1+\omega)}
{4[1-\frac{1}{2} C(1+\omega)][1-C(1+\omega)+\omega C^2]}\,\,\,\,,
\label{27}
\end{eqnarray}\\
where $C=\beta (1-q)$. Taking the limit $T\rightarrow 0$ and
focusing attention on pure states, where
$m_{\mu}=m\delta_{\mu\lambda}$, for some $\lambda$, we can do the
integrations and the averages and follow the usual procedure
\cite{AGS87} to reduce Eqs.(25)-(27) to a single equation, now in
terms of the scaled overlap
\begin{equation}
x=m/\sqrt{2\alpha [\tilde r+J_0^2(1-c)/c]}\,\,\,, \label{28}
\end{equation}
where $\tilde r$ is given by Eq. (27) and $q\rightarrow 1$ in the
limit $T\rightarrow 0$. We also have

\begin{equation}
C\sim \Big[\frac{2}{\pi\alpha [\tilde
r+J_0^2(1-c)/c]}\Big]^{\frac{1}{2}}\exp(-x^2) \label{29}
\end{equation}
and
\begin{equation}
x\sqrt{\alpha} =
\frac{erf(x)\,[A(x,\omega)]^{\frac{1}{2}}}{[(1+\omega^2)\,erf^2(x)\,B(x,\omega)
+\frac{1}{2}\,(1+\omega)^2\,\frac{(1-c)}{c}\,A(x,\omega)]^{\frac{1}{2}}}
 \,\,\,\,, \label{30}
\end{equation}
where
\begin{eqnarray}
A(x,\omega)&=&[erf(x)-\frac{2x}{\sqrt{\pi}}\exp(-x^2)][erf(x)
-\frac{(1+\omega)x}{\sqrt{\pi}}\exp(-x^2)] \nonumber \\
&\times&[erf(x)-\frac{2\omega x}{\sqrt{\pi}}\exp(-x^2)]\,\,\,\,, \label{31}\\
B(x, \omega)&=&erf(x)-\frac{(\omega + \omega^2)}{(1+\omega^2)}\,
\frac{2\,x}{\sqrt{\pi}}\exp(-x^2)  \label{32}\,\,\,.
\end{eqnarray}\\
When $c=1$ one recovers the relationship for the fully connected
dual model \cite{CV96}, and when $\omega=-1$ ($J_0=0$) or
$\omega=1$ ($J=0$) one gets back the result for the layered
network \cite{DKM89} or the result for the diluted recurrent
network \cite{CN92}, respectively.

The numerical solution of Eqs.(30)-(32) yields the critical
storage capacity $\alpha_c(\omega, c)$, for a given $\omega$ and
$c$, as the value of $\alpha$ where the solution with finite
overlap (non-zero $x$) disappears. The result is shown in Fig. 2
for various values of $c$. When $\omega=1$ (the purely recurrent
network) one recovers the $\alpha_c=0.138$ for the fully connected
network with $c=1$ \cite{AGS87}. On the other hand, when
$\omega=-1$ (the purely layered network) $\alpha_c=0.269$ for any
value of $c$, since the dilution is only in the recurrent part
\cite{DKM89}. There is a maximum value of $\alpha_c\sim 0.317$ for
$\omega\sim -0.12$, when $c=1$, in consistency with a previous
result \cite{CV96}, and the maximum decreases and shifts towards
the behavior of the layered network with increasing dilution. It
has also been checked that although $\alpha_c$ decreases with
increasing dilution, the ratio $\alpha_c/c=0.629$, as one would
expect in the extreme diluted limit $c\rightarrow 0$ for
$\omega=1$.

Although $\alpha_c$ is a decreasing function of $c$ for all values
of $\omega$, as one would expect, the network still has a
relatively large storage capacity for a moderate amount of
dilution and a balanced ratio of interaction strengths. For
instance, for $c=0.1$ and $\omega\approx -0.20$ ($J/J_0\approx
1.5$), $\alpha_c$ is still the same as that for the fully
connected network. Thus, the synaptic interactions between layers
do not need to be much stronger than the interactions within the
layers for a typical good performance with a moderately finite
degree of dilution. As pointed out in the introduction, this may
be of use for biological networks.\\
\mbox{} \\
\noindent 4.2 {\it{Multiple phase transitions in the first layers}}\\
\mbox{} \\
We consider now the operation of the first two layers and we study
the effects of dilution on multiple phase transitions that already
appear at the junction between the first and the second layer
\cite{CV96}. Even limiting the study to pure states as we do here,
with $m_{\mu}=m\delta_{\mu\lambda}$ for some $\lambda$, the
overlap is layer dependent and we denote it by $m$ and
$m^{\prime}$ for the first and the second layer, respectively.

We deal first with the clamped operation in the first layer, with
a given overlap vector $\bm$, and take the $T\rightarrow 0$ limit
in Eqs.(18),(19) to obtain $m^{\prime}=erf(y)$, where
\begin{equation}
y=\frac{J_0m^{\prime}+Jm}{\sqrt{2\alpha[\tilde
r^{\prime}+J_0^2(1-c)/c]}}\,\,\,, \label{33}
\end{equation}
is the appropriate scaled overlap and $\tilde r^{\prime}$ is given
by Eq.(21). In that limit, $q^{\prime}\rightarrow 1$ and
$C^{\prime}=\beta(1-q^{\prime})$ becomes
\begin{equation}
C^{\prime}\sim \Big[\frac{2}{\pi\alpha [\tilde
r^{\prime}+J_0^2(1-c)/c]}\Big]^{\frac{1}{2}}\exp (-y^2)\,\,\,.
\label{34}
\end{equation}
When this is used we obtain the non-linear equation for $y$,
\begin{equation}
G(y,\alpha,m)\Big[F(y)+m\frac{1-\omega}{1+\omega}\Big]=y\sqrt{2\alpha}
\Big[1+\rho
(\frac{1-\omega}{1+\omega})^2\Big]^{\frac{1}{2}}\,\,\,, \label{35}
\end{equation}
in which
\begin{eqnarray}
F(y)&=&erf(y)-\frac{2y}{\sqrt{\pi}}e^{-y^2}\,\,\,, \label{36}\\
G^2(y,\alpha,
m)&=&1-(\frac{1-c}{c})\Big[\frac{y\sqrt{2\alpha}}{erf(y)
+m\frac{1-\omega}{1+\omega}}\Big]^2\,\,\, \label{37}
\end{eqnarray}
and $\rho=1$. Solution of Eq.(35) with Eqs.(36) and (37) yields
the overlap $m^{\prime}$ for the second layer and the phase
diagrams presented below.

We consider next the case of free relaxation of the first layer,
and restrict ourselves again to pure states. First, Eqs.(22)-(24)
have to be solved for the scaled overlap in the limit
$T\rightarrow 0$ and when the solution is taken as an input in
Eqs.(18)-(20), in the same limit, one obtains a new overlap
$m^{\prime}$ on the second layer for this mode of operation. Thus,
we find first $m=erf(w)$, where
\begin{equation}
w=J_0m/\sqrt{2\alpha[\tilde r+J_0^2(1-c)/c]}\,\,\, \label{38}
\end{equation}\\
and $C=\beta(1-q)$ is now given by Eq.(29) with $w$ in place of
$x$ and $F(w)$ is given by the same expression as in Eq.(36). We
find the equation
\begin{equation}
F(w)=\frac{w\sqrt{2\alpha}}{G(w,\alpha,0)} \label{39}
\end{equation}
that yields the scaled overlap $w$ for free relaxation of the
first layer. Following similar steps as for the clamped operation,
with the full expression for $\tilde r^{\prime}$ in Eq.(20), in
place of Eq.(21), we find formally the result given by Eq.(35) but
now with
\begin{equation}
\rho=[\frac{m}{F(w)}]^2\,\,\,. \label{40}
\end{equation}
When $c=1$, $G(y,\alpha,m)=1$ for any $m$, and we recover the
results for the fully connected network.

It can be seen from the above equations that, as long as the
dilution remains finite, there is a clear distinction between the
two modes of operation, as in the fully connected network
\cite{CV96}. Already when $m=0$, in the case of a clamped input
with $\rho=1$, Eqs.(35)-(37) yield a critical $\alpha_c$ for a
bifurcation of a solution with $y\neq 0$ (meaning $m^{\prime}\neq
0$), for any finite $c$. In contrast, one finds that $\rho=\infty$
for free relaxation of the first layer and in that case the only
solution is $y=0$, that is, $m^{\prime}=0$.

To obtain all the solutions that appear for general values of $m$
(in the case of clamped input) and $c$, we look for the
bifurcations from Eq.(35) that are given by this equation and its
derivative with respect to $y$ which have to be solved
simultaneously. This can be done either for the clamped input,
with $\rho=1$, or for the case of a free relaxation of the first
layer, with $\rho$ given by Eq.(40). The outcome for the case of a
clamped input is shown in Fig. 3 for $m=0$ and in Fig. 4 for
$m=1$, in both cases for various values of $c$, together with the
saturation transition for the infinite chain in each case, for
reference. This is not the saturation transition for the chain of
two layers we are discussing here, except for $\omega=1$, as will
be seen below.

The phase diagram for the two-layer chain when $m=0$ has two
regions. In region I, everywhere above the solid phase boundary
$m^{\prime}=0$ is the only solution. In region II, below the solid
line, there are three stable solutions: $m^{\prime}_1\sim 1$,
$m^{\prime}_2\sim -1$ and $m^{\prime}_3=0$, such that
$|m^{\prime}_1|\neq |m^{\prime}_2|$, except for $\omega=1$ where
$|m^{\prime}_1|=|m^{\prime}_2|$. The phase boundary of multiple
solutions of region II meets the saturation transition of the
infinitely long chain at $\omega=1$. This is the place where the
model becomes a set of purely recurrent networks, with no
interaction between layers and, hence, the stationary states of
the second layer are already those of the infinite chain. The
solution with overlap zero and $q^{\prime}=1$, above and below the
phase boundary, describes a spin-glass state.

In the case of an input $m=1.0$, one obtains two further regions
of coexisting stable states shown in Fig. 4. Now in region II one
has two solutions, $m^{\prime}_1\sim 1$ and $m^{\prime}_2\sim -1$,
such that $|m^{\prime}_1|\neq |m^{\prime}_2|$. In region III there
are two stable retrieval solutions with $m^{\prime}_1\sim 1$,
$m^{\prime}_2$ smaller than one. In region IV there are three
stable retrieval solutions $m^{\prime}_1\sim 1$, $m^{\prime}_2\sim
-1$ and $m^{\prime}_3$ small and positive, such that
$|m^{\prime}_1|\neq |m^{\prime}_2|$, except for $\omega=1$ where
$|m^{\prime}_1|=|m^{\prime}_2|$. The small solution $m^{\prime}_3$
vanishes precisely at $\omega=1$, leaving the known results for
the purely recurrent network, regardless of dilution. The phase
boundaries that go up to $\omega=1$ again meet there the
saturation transition of the infinitely long chain.

Again, all three regions of retrieval are reduced with increasing
dilution, particularly regions III and IV, but they are still
there for a finite dilution of $c=0.1$. In the case of extreme
dilution only region II is left over a tiny part of the phase
diagram in terms of $\alpha=p/N$. But in this limit it is more
appropriate to consider $\alpha^{\prime}=\alpha/c$, which remains
finite. Similar results are obtained for other finite values of
the fixed input overlap $m$. We also analyzed the effects of free
relaxation of the first layer on the performance of the second
layer and found qualitatively similar results to those described
here for the case of clamped input.

As one can see, the presence of these regions and the coexistence
of various retrieval states is a feature of the competition
between the recurrent and the layer information processing, which
is considerably diminished with increasing dilution. When $c=1$ we
recover the earlier results for the fully connected network
\cite{CV96}.\\
\mbox{} \\
\noindent 4.3 {\it{The replica-symmetry breaking (RSB) transition}}\\
\mbox{} \\
We look here for the effect of synaptic dilution on the de
Almeida-Thouless (AT) transition \cite{AT78} where the replica
symmetric solution for the saddle point equations ceases to be
valid at the AT line. There is RSB at sufficiently low $T$ in the
fully recurrent network, that is when $\omega=1$, and there is no
RSB in the purely layered network with $\omega=-1$. In both, the
fully connected and the diluted dual model RSB appears as soon as
$\omega>-1$, that is already for an arbitrarily small recurrent
interaction \cite{CV96}. Although the AT line is a boundary in the
$(T,\alpha)$ plane and our work, as well as that in that ref.
[11], are restricted to $T=0$ one may get an idea of the size of
the RSB region by looking at the deformation of that boundary with
a decrease of $\omega$ towards $\omega=-1$, and that is what we do
next.

The AT line in the present model is obtained considering the
fluctuations around the replica symmetric forms for the spin glass
order parameter
$q_{\alpha\beta}=\langle\sigma_{i\alpha}\sigma_{i\beta}\rangle$
and the auxiliary parameter $\hat q_{\alpha\beta}$ given by the
right-hand-side of Eq.(15) for $\alpha \neq \beta$. Proceeding in
the usual way on the free energy prior to the assumption of
replica symmetry we obtain the AT line for the diluted dual model,
\begin{eqnarray}
&1&=\alpha\beta^2\{\frac{J_0^2}{[1-\beta
J_0(1-q)]^2}+J_0^2(1-c)/c\}\Big\langle\int Dz\,\,\nonumber\\
&\times&\cosh^{-4}\beta\Big\{\bxi.(J_0\bm+J\btm)+z\sqrt{\alpha
[\tilde r+J_0^2(1-c)q/c]}\,\Big\}\Big\rangle_{\bxi}\,\,\,.
\label{41}
\end{eqnarray}
We recover the expression for the purely recurrent network when
$J_0=1$ and $J=0$ \cite{CN92} and that for the fully connected
dual model when $c=1$ \cite{CV96}. The replica symmetric solution
Eqs.(18)-(20) is stable when the right-hand side of Eq.(41) is
less than one. This corresponds to the regions above the AT lines
shown in Fig. 5 for three different values of $\omega$ when $c=1$,
$c=0.1$ and $c=0.001$. It can be seen that the domains of RSB
below the AT lines continue to become smaller with a decrease of
$\omega$ (that is with reduced recurrent interactions) towards
$\omega=-1$ where the AT line coincides with the $T=0$ axis as in
the fully connected network model \cite{CV96}. The effect of
(finite) dilution is to increase the region of RSB with a shift of
the AT lines towards smaller values of $\alpha=p/N$. Nevertheless,
the effects of RSB are smaller than in the purely recurrent
network justifying the analysis in this work based on the
assumption of replica symmetry.

\section{Conclusions}

We studied in this work the effects of finite symmetric dilution
on the performance of a dual model that combines information
processing of recurrent and feed-forward networks. The model
consists of a feed-forward chain of recurrent networks and the
dilution is in the symmetric synaptic interactions of the
recurrent layers. Our analysis extends the original work of Coolen
and Viana with symmetric synaptic interactions between units
within the layers and non-symmetric interactions between units in
consecutive layers. The competition between these interactions is
responsible for the behavior shown in the phase diagrams. On one
hand, a small inter-layer feed-forward interaction between layers
produces the multiple transitions that are found either in the
case of clamped operation with finite $m$ or in the case of free
relaxation. In the other extreme, an infinitesimal symmetric
recurrent interaction already yields a RSB transition.

A stationary state replica analysis was carried out in this work
assuming a frozen-in dilution of synaptic interactions in order to
study the saturation transition in infinitely long chains and the
multiple transitions that already appear at the junction between
the first and second layer. We found that finite synaptic dilution
produces a gradual change of the performance of the model reducing
all the regions of stable retrieval states, in particular those
where multiple solutions appear due to the competition between
recurrent and feed-forward processing. One would expect that the
study of further initial layers should yield a performance closer
to that of long chains.

Although the performance of the dual model is reduced by synaptic
dilution, we showed that there can still be a considerable output
in a long chain of layers in the case of a finite amount of
dilution of the synapses within the layers and pointed out that
this feature could be relevant for dual models of biological
networks.

There are several extensions that one may conceive of the work
presented in this paper. It would be interesting to consider the
effects of non-linear synapses of a general form \cite{So86} and
one could also think of relaxing the symmetry of the synaptic
interactions within the layers. A dynamical approach would be
necessary in that case.\\
\mbox{} \\
\noindent{\bf Acknowledgments}\\
\mbox{} \\
One of the authors (WKT) thanks Desir\'e Boll\'e for informative
discussions and for the kind hospitality at the Institute of
Theoretical Physics of the Catholic University of Leuven, Belgium.
The work of the same author was financially supported, in part, by
CNPq (Conselho Nacional de Desenvolvimento Cient\'{\i}fico e
Tecnol\'ogico), Brazil. Grants from CNPq and FAPERGS
(Funda\c{c}\~ao de Amparo \`a Pesquisa do Estado de Rio Grande do
Sul), Brazil, to the same author are gratefully acknowledged. F.
L. Metz acknowledges a graduate student fellowship from CNPq.

\vspace{0.5cm}
\noindent{\bf Figures}
\begin{figure}[ht]
\begin{center}
\centering
\includegraphics[width=8cm,height=7cm]{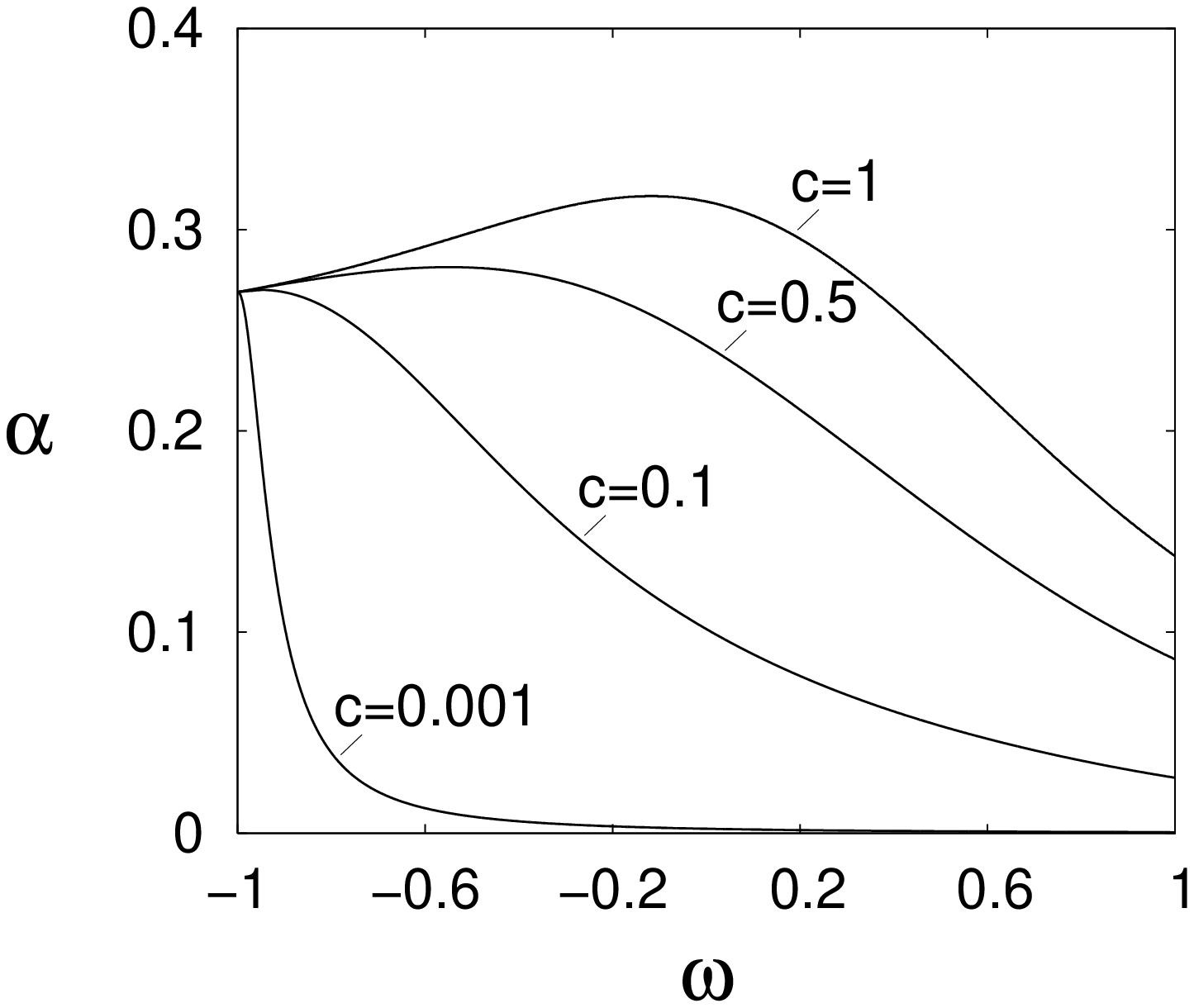}
\end{center}
\caption{  Critical storage capacity $\alpha_c$ for an infinitely
long chain with various degrees of dilution, as indicated. The
fully connected network, with $c=1$, is also shown for reference.
Increasing dilution is described by a decreasing value of $c$.}
\label{fig2}
\end{figure}

\begin{figure}[ht]
\begin{center}
\centering
\includegraphics[width=8cm,height=16cm]{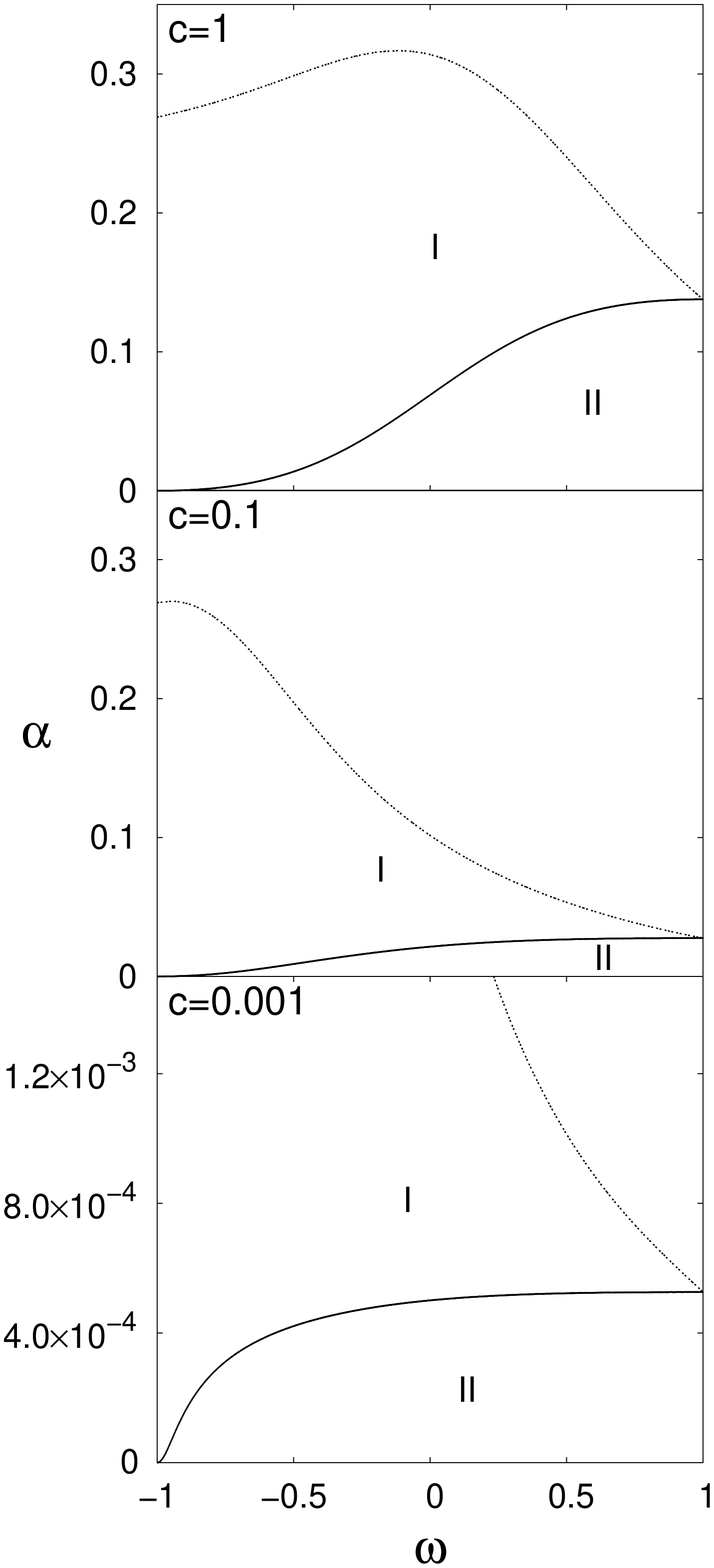}
\end{center}
\caption{ Phase diagram for the two-layer network under clamped
operation with overlap $m=0$ in the first layer for various
degrees of dilution, as indicated. The saturation transition (in
dotted lines) is shown for reference and the phases I and II are
described in the text.} \label{fig3}
\end{figure}

\begin{figure}[ht]
\begin{center}
\centering
\includegraphics[width=8cm,height=16cm]{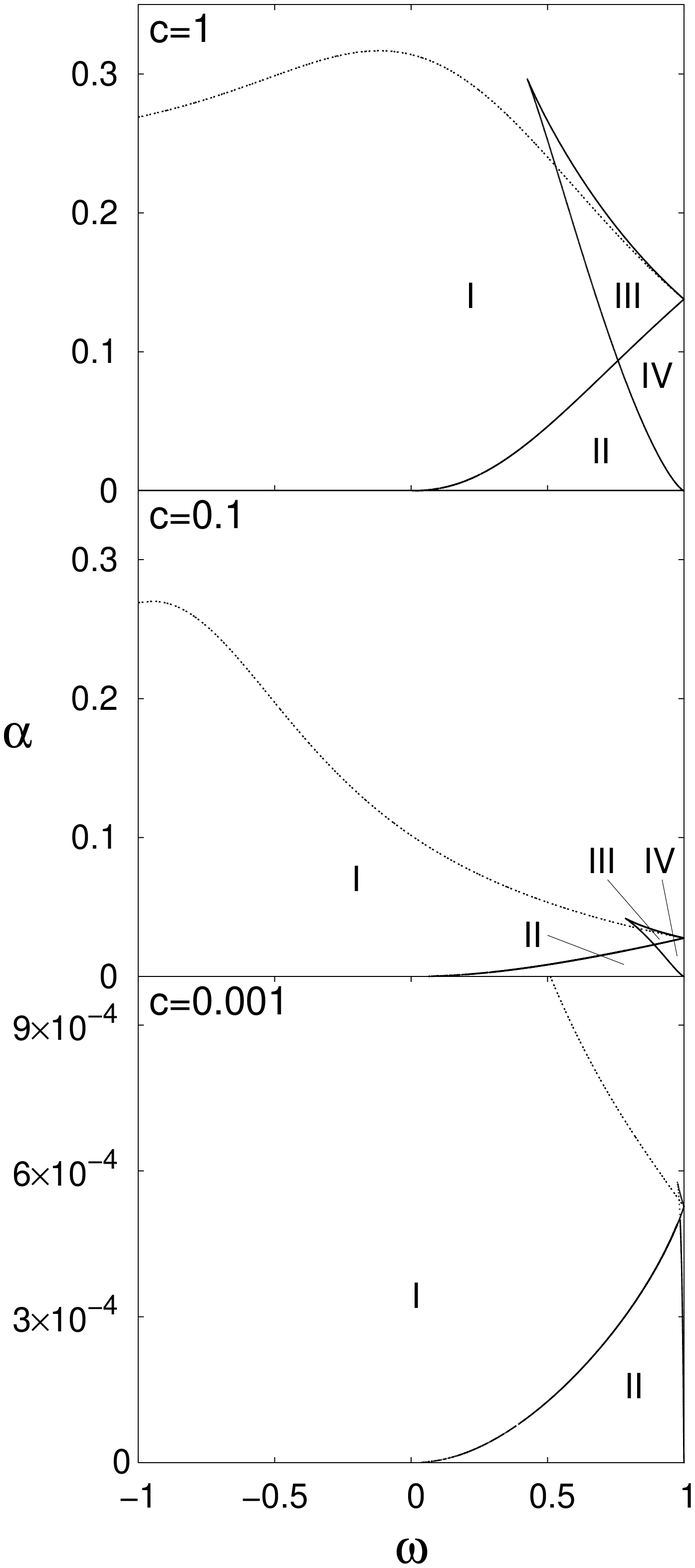}
\end{center}
\caption{ Phase diagram for the two-layer network under clamped
operation with overlap $m=1$ in the first layer for various
degrees of dilution, as indicated. The saturation transition (in
dotted lines) is shown for reference and the phases I-IV are
described in the text.} \label{fig4}
\end{figure}

\begin{figure}[ht]
\begin{center}
\centering
\includegraphics[width=8cm,height=13cm]{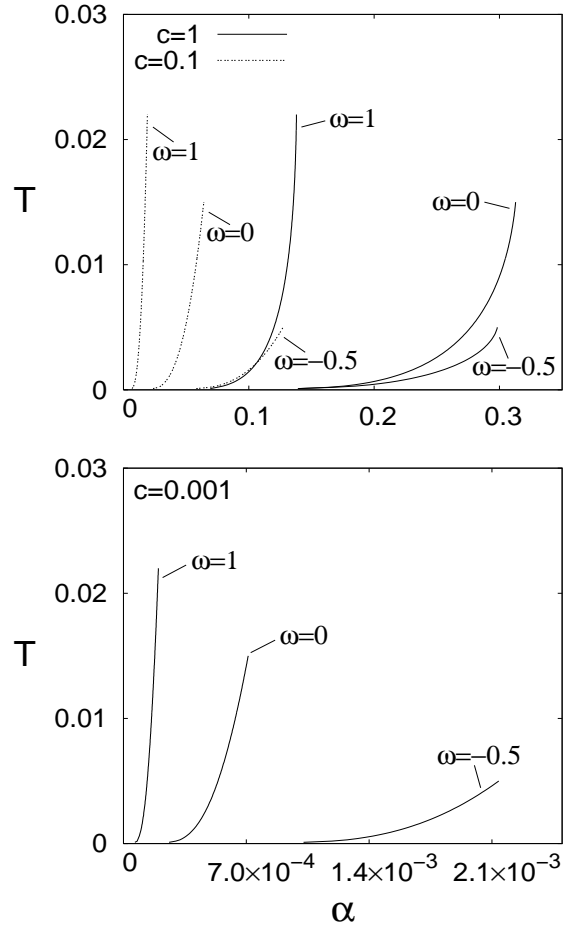}
\end{center}
\caption{ The replica-symmetry breaking transition (AT line) for
$\omega=1.0,\,0,\,-0.5$ and various degrees of dilution. For
$\omega=-1$ (purely layered network) the AT line coincides with
the $T=0$ line.} \label{fig5}
\end{figure}

\end{document}